# Boundary integral equation method for resonances in gradient index cavities designed by conformal transformation optics

**Jung-Wan Ryu**[1,†], **Jinhang Cho**[2,†], **Soo-Young Lee**[3,‡], **Yushin Kim**[4], **Sang-Jun Park**[3], **Sunghwan Rim**[2], **Muhan Choi**[2,3,*] **& Inbo Kim**[2,*]

[1] Center for Theoretical Physics of Complex Systems, Institute for Basic Science (IBS), Daejeon 34126, Republic of Korea
[2] Digital Technology Research Center, Kyungpook National University, Daegu 41566, Republic of Korea
[3] School of Electronics Engineering, Kyungpook National University, Daegu 41566, Republic of Korea
[4] KU-KIST Graduate School of Converging Science and Technology, Korea University, Seoul 02841, Republic of Korea
† these authors contributed equally to this work
‡ deceased
* mhchoi@ee.knu.ac.kr; ibkim@ee.knu.ac.kr ; co-corresponding

## ABSTRACT

In the case of two-dimensional gradient index cavities designed by the conformal transformation optics, we propose a boundary integral equation method for the calculation of resonant mode functions by employing a fictitious space which is reciprocally equivalent to the physical space. Using the Green's function of the interior region of the uniform index cavity in the fictitious space, resonant mode functions and their far-field distributions in the physical space can be obtained. As a verification, resonant modes in limaçon-shaped transformation cavities were calculated and mode patterns and far-field intensity distributions were compared with those of the same modes obtained from the finite element method.

## 1. Introduction

Recently, transformation optics (TO) has been attracting great interest because it provides the theoretical basis of incredible optical gear such as optical invisibility cloak [1-4] and realization of celestial objects such as optical black holes in laboratory [5]. Also, beyond optics it is widely applicable to acoustic waves [6, 7], elastic waves [8, 9], and seismic waves [10], etc. To date, many studies in TO have mainly been focused on devices enabling the control of light path, such as invisibility cloaks, flat Luneburg lenses [11, 12], and waveguide bends [13], to name a few. In addition to these applications of TO, recently Kim *et al*. have suggested that TO can be exploited to the design of two-dimensional (2D) optical dielectric cavities [14]. They have designed a new type of gradient index (GRIN) resonator called transformation cavity (TC) by using conformal TO and showed that high-*Q* anisotropic whispering gallery modes (WGMs) exhibiting desired directional emissions can be formed. Basically, WGMs are resonances with very long lifetime (which means very high Q-factor) due to the total internal reflection (TIR) of light around the rim of the cavities and they play a central role in photonic devices such as lasers and sensors.

To study the properties of the resonant modes formed in these TCs, numerical calculation of the spatial configuration of electromagnetic field, i.e., mode functions are indispensable. Typically, domain type methods such as finite element method (FEM) or finite difference method (FDM) are generally used for this inhomogeneous potential problem, while boundary element method (BEM) is not generally applicable because the Green's function of the interior region of the GRIN cavities is not known. But, boundary-only methods such as dual reciprocity method (DRM) [15] and analog equation method (AEM) [16] which have been developed from the pure BEM can be used for this kind of problems with inhomogeneous material properties. These boundary-only formalism commonly uses internal nodes (collocation points) in the inhomogeneous potential region and convert domain integrals into boundary integrals, which is a rather complicated and cumbersome procedure and there is no report, as far as we know, that these methods have been employed for the optical resonant mode calculation of the 2D GRIN dielectric cavities.

On the other hand, in case of 2D dielectric cavities with uniform refractive indices, BEM has been widely used because of its advantages over other methods [17-19]. In particular, BEM works well for strongly deformed geometries from the circular shape and for the calculation of highly-excited resonant states (short wavelength regime) since it uses only the field information at the discretized cavity boundary, while FEM which is based on the 2D discretized space causes a heavy computational load [20, 21]. In addition, because an outgoing-wave boundary condition at infinity is naturally built into Green's function in BEM, one does not need to artificially construct an absorbing boundary environment at the outermost computational region such as the perfectly matched layer (PML), which is required in FEM for the truncation of the unlimited outside region [22]. Thus, BEM has been extensively used in calculating resonant modes in various 2D uniform index dielectric cavities, e.g., deformed cavities [23, 24], coupled cavities [25], and annular cavities [26].

In this paper, we show that conventional BEM can still be used to calculate resonant modes of TCs with a spatially-varying refractive index profile determined by an optical conformal mapping. To this end, we introduce a reciprocal virtual (RV) space which is related to the physical space by the inverse conformal mapping. The cavity in the RV space has a uniform refractive index and there one can build a boundary integral equation (BIE) for the interior region of the cavity.

The paper is organized as follows. The wave equations in a wholly transformed (WT) space are derived from Maxwell's equations in an original virtual (OV) space in section 2. In section 3, we formulate BEM for the resonant modes of TCs by using the RV space. In section 4, we demonstrate the validity of our BEM with examples of limaçon-shaped TCs. Finally, we summarize our work in section 5.

## 2. GRIN cavities designed by conformal transformation optics

The theories of TO have been independently proposed by Ulf Leonhardt (conformal TO with Helmholtz equation in complex plane) [1] and Sir John Pendry (general TO with Maxwell's equations) [2] as design methodologies of their optical invisibility cloaks. The general TO is based on the form invariance of Maxwell's equations under a general coordinate transformation; as the consequences of the form invariance, electric and magnetic fields are renormalized and the constitutive parameters (i.e., the electric permittivity $\varepsilon$ and the magnetic permeability $\mu$) are changed to anisotropic tensor quantities [2, 27].

Here, we will derive the conformal TO from the general TO and then obtain the wave equation for infinitely-long cylindrical dielectric cavity in a WT space. We start from an OV space described by Cartesian coordinates $(u, v, w)$, where the Maxwell's equations in frequency domain ($\sim e^{-i\omega t}$) without sources or currents in linear isotropic dielectric media are written as [13]

$$\nabla \times \mathbf{E} = i\omega\mu \mathbf{H}, \quad \nabla \times \mathbf{H} = -i\omega\varepsilon \mathbf{E}. \tag{2.1}$$

In Eq. (2.1), $\mathbf{E}$ and $\mathbf{H}$ are time-independent complex electric and magnetic field, respectively; the magnetic permeability $\mu$ is given by $\mu_0\mu_r$, where $\mu_0$ and $\mu_r$ are the vacuum permeability and the relative permeability, respectively; and the electric permittivity $\varepsilon$ is given by $\varepsilon_0\varepsilon_r$, where $\varepsilon_0$ and $\varepsilon_r$ are the vacuum permittivity and the relative permittivity, respectively.

Under a general coordinate transformation from the OV space to a WT space described by coordinates $(x, y, z)$, Maxwell's equations that keep their forms invariant in the WT space transforms as

$$\nabla' \times \mathbf{E}' = i\omega\boldsymbol{\mu}' \mathbf{H}', \quad \nabla' \times \mathbf{H}' = -i\omega\boldsymbol{\varepsilon}' \mathbf{E}', \tag{2.2a}$$

where

$$\mathbf{E}' = (\Lambda^T)^{-1} \mathbf{E}, \quad \mathbf{H}' = (\Lambda^T)^{-1} \mathbf{H}, \quad \nabla' = (\Lambda^T)^{-1} \nabla. \tag{2.2b}$$

The relation of the constitutive parameters between the OV space and the WT space are given by

$$\boldsymbol{\varepsilon}' = \frac{\varepsilon \, \Lambda \Lambda^T}{|\det \Lambda|} , \qquad \boldsymbol{\mu}' = \frac{\mu \, \Lambda \Lambda^T}{|\det \Lambda|} , \qquad (2.3)$$

where $\Lambda$ is the Jacobian matrix of the transformation from the OV space to the WT space, and is given by

$$\Lambda = \begin{bmatrix} \frac{\partial x}{\partial u} & \frac{\partial x}{\partial v} & \frac{\partial x}{\partial w} \\ \frac{\partial y}{\partial u} & \frac{\partial y}{\partial v} & \frac{\partial y}{\partial w} \\ \frac{\partial z}{\partial u} & \frac{\partial z}{\partial v} & \frac{\partial z}{\partial w} \end{bmatrix} . \qquad (2.4)$$

According to the so-called material interpretation [28] or manipulation-of-material viewpoint [29], one can think that the effect of coordinate transformation is encoded in the redefined fields and anisotropic constitutive parameters in unchanged flat Cartesian space. In conformal TO, the coordinate transformation is given by an analytic function $\zeta = f(\eta) = x(u,v) + iy(u,v)$ of a complex variable $\eta = u + iv$ with $z = w$, so the Jacobian matrix $\Lambda$ is reduced to a block diagonal form,

$$\Lambda = \begin{pmatrix} \frac{\partial x}{\partial u} & \frac{\partial x}{\partial v} & 0 \\ \frac{\partial y}{\partial u} & \frac{\partial y}{\partial v} & 0 \\ 0 & 0 & 1 \end{pmatrix} , \qquad (2.5)$$

because, $\frac{\partial x}{\partial w} = \frac{\partial y}{\partial w} = 0$, and $\frac{\partial z}{\partial u} = \frac{\partial z}{\partial v} = 0$. The analytic function $\zeta = f(\eta)$ always produces a conformal coordinate transformation, which satisfies the Cauchy–Riemann equations,

$$\frac{\partial x}{\partial u} = \frac{\partial y}{\partial v}, \qquad \frac{\partial y}{\partial u} = -\frac{\partial x}{\partial v} . \qquad (2.6)$$

Due to these relations and orientation-preserving property of conformal mappings (i.e., $\det \Lambda > 0$), the constitutive parameters in the WT space turn out to have a simple diagonal form with the scale factor of a conformal mapping squared as the third diagonal element as follows [30],

$$\boldsymbol{\varepsilon}' = \varepsilon \begin{pmatrix} 1 & 0 & 0 \\ 0 & 1 & 0 \\ 0 & 0 & \frac{1}{\det \Lambda} \end{pmatrix}, \qquad \boldsymbol{\mu}' = \mu \begin{pmatrix} 1 & 0 & 0 \\ 0 & 1 & 0 \\ 0 & 0 & \frac{1}{\det \Lambda} \end{pmatrix} . \qquad (2.7)$$

For this case of conformal TO, let us consider an infinitely-long cylindrical dielectric cavity with a circular cross section in the OV space (Figure 1(a)) from which cylindrical cavity with a deformed cross section in a WT space (Figure 1(b)) can be obtained by a conformal mapping, taking $z$-axis and $w$-axis along the length of the cylinder [14, 19]. For the transverse magnetic (TM) polarized mode of the cylindrical cavity, only the $z$-component, $E_z'$ among the components of electric field is non-zero, i.e., $\mathbf{E}' = E_z' \hat{z}$, $\mathbf{H}' \cdot \hat{z} = 0$, so Eq. (2.2a) becomes

$$\boldsymbol{\nabla}' \times \mathbf{E}' = i\omega\mu \, \mathbf{H}', \qquad \boldsymbol{\nabla}' \times \mathbf{H}' = -i\omega\boldsymbol{\varepsilon}' \, \mathbf{E}'. \qquad (2.8)$$

From these equations, following wave equation for the TM mode can be obtained by using another Maxwell equation, $\boldsymbol{\nabla}' \cdot (\boldsymbol{\varepsilon}' \mathbf{E}') = 0$,

$$[\Delta' + n'^2(\mathbf{r})k^2] \, E_z'(x,y) = 0 , \qquad (2.9)$$

where $\Delta' \equiv \partial_x^2 + \partial_y^2$, and the spatially-varying refractive index $n'(\mathbf{r})$ in the WT space is given by

$$n'(\mathbf{r}) = \begin{cases} \dfrac{\sqrt{\varepsilon_r \mu_r}}{\sqrt{\det \Lambda}} = n_0 \left|\dfrac{d\zeta}{d\eta}\right|^{-1}, & \mathbf{r} \in \Omega_{in} \\ \dfrac{1}{\sqrt{\det \Lambda}} = \left|\dfrac{d\zeta}{d\eta}\right|^{-1}, & \mathbf{r} \in \Omega_{ex}, \end{cases} \quad (2.10)$$

where $n_0$ is the refractive index of the cavity in an OV space which is equal to $\sqrt{\varepsilon_r \mu_r}$; the free space wave number is $k = \omega/c$, where $c$ is the speed of light in vacuum given by $1/\sqrt{\varepsilon_0 \mu_0}$; $\Omega_{in(ex)}$ denotes the interior (exterior) region of the cavity. Equation (2.9) can be also obtained starting from 2D Helmholtz equation under conformal mapping, which is a well-known result in waveguide theory [31]. At the dielectric interface, both $E'_z(x, y)$ and its normal derivative $\partial_\perp E'_z(x, y)$ are continuous across the cavity boundary [32],

$$E'_z|_{in} = E'_z|_{ex}, \qquad \partial_\perp E'_z|_{in} = \partial_\perp E'_z|_{ex} \quad (2.11).$$

The normal derivative is defined as $\partial_\perp \equiv \mathbf{p}(\mathbf{r}) \cdot \nabla|_\mathbf{r}$ where $\mathbf{p}(\mathbf{r})$ is the outward normal unit vector at point $\mathbf{r}$ on the boundary curve $\Gamma$ of the interior (exterior) region of the cavity. The TM modes of a cylindrical cavity in the WT space are supported by the refractive index $n'$ originating from the anisotropic electric permittivity, $\boldsymbol{\varepsilon}'$, not from the anisotropic magnetic permeability, $\boldsymbol{\mu}'$ as can be seen from Eq. (2.8). It is noted that the electric field wave function $E'_z(x, y)$ in the WT space and its counterpart function $E_w(u, v)$ in the OV space are the same, as can be shown by Eq. (2.2b) and Eq. (2.5).

For the transverse electric (TE) polarized mode of a cylindrical cavity in a WT space, among the components of the magnetic field, only the $z$-component, $H'_z$ is non-zero, i.e., $\mathbf{H}' = H'_z \hat{z}$, $\mathbf{E}' \cdot \hat{z} = 0$ then Eq. (2.2a) becomes

$$\boldsymbol{\nabla}' \times \mathbf{E}' = i\omega \boldsymbol{\mu}' \mathbf{H}', \qquad \boldsymbol{\nabla}' \times \mathbf{H}' = -i\omega\varepsilon \mathbf{E}'. \quad (2.12)$$

From these equations, following wave equation for the TE mode can be obtained by using another Maxwell equation, $\boldsymbol{\nabla}' \cdot (\boldsymbol{\mu}' \mathbf{H}') = 0$,

$$[\Delta' + n'^2(\mathbf{r})k^2] H'_z(x, y) = 0, \quad (2.13)$$

where the spatially-varying refractive index $n'(\mathbf{r})$ in the WT space is the same as the one in Eq. (2.10). At the dielectric interface, $H'_z(x, y)$ and instead of its normal derivative, $(1/n'^2)\partial_\perp H'_z(x, y)$ are continuous across the cavity boundary [32],

$$H'_z|_{in} = H'_z|_{ex}, \qquad \left.\dfrac{\partial_\perp H'_z}{n'^2}\right|_{in} = \left.\dfrac{\partial_\perp H'_z}{n'^2}\right|_{ex}. \quad (2.14)$$

TE modes of a cylindrical cavity in the WT space are supported with the refractive index $n'$ originating from the anisotropic magnetic permeability, $\boldsymbol{\mu}'$, not from the anisotropic electric permittivity, $\boldsymbol{\varepsilon}'$, as one can see from Eq. (2.12). It is also noted that the magnetic field wave functions $H'_z(x, y)$ in the WT space and their counterpart functions $H_w(u, v)$ in the OV space are the same, as can be shown by Eq. (2.2b) and Eq. (2.5).

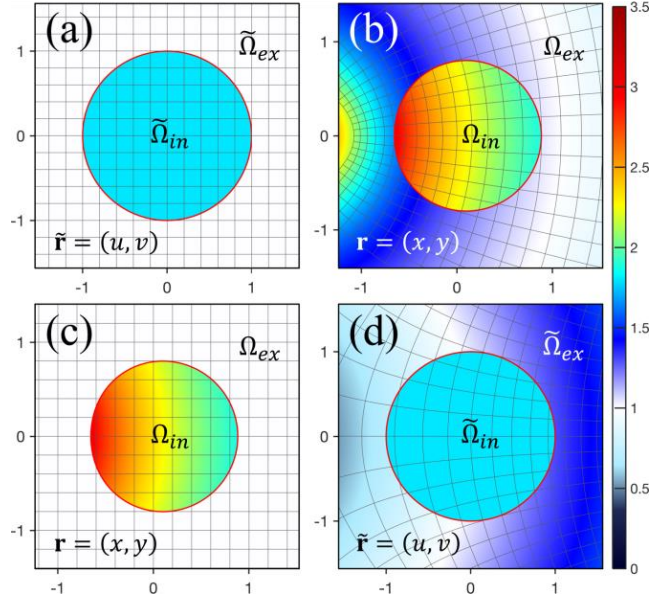

**Figure 1.** Schematic illustration for (a) a circular cavity in the OV space ($\eta = u + iv$), (b) a limaçon-shaped cavity in the WT space ($\zeta = x + iy$), (c) the corresponding limaçon-shaped TC in the physical space ($\zeta = x + iy$), and (d) the inverse transformed circular cavity in the RV space ($\eta = u + iv$) obtained by conformal mapping described in section 4. (Straight gray lines in (a) and (c) are grids of coordinates. Curved gray lines in (b) and (d) are the transformed image of the straight grid lines in (a) and (c) by the conformal mapping. Scaled color represents refractive index.)

## 3. Boundary element method for resonant modes in transformation cavities

The GRIN cavity in WT space obtained from the homogeneous disk cavity in the OV space by an optical conformal mapping is surrounded by inhomogeneous material that fills infinitely extended outer region, which is an unphysical situation. Thus, we replace the outside inhomogeneous index with 1, i.e., the refractive index of air, as shown in Figure 1(c) and then call the same cavity a transformation cavity (TC) in the physical space, which has gradually varying refractive index profile, $n(x, y)$ in the interior region of the cavity according to Eq. (2.10) [14]. In other words, conformal mapping is only applied to the interior region ($\widetilde{\Omega}_{in}$) of the cavity in the OV space to obtain a TC in the physical space. Hereafter, primed quantities and operators in the WT space used in the previous section will be replaced with unprimed ones in the physical space for convenience.

Now let us formulate a BEM for optical resonant modes in a TC with inhomogeneous refractive index in physical space. We have to solve the following scalar wave equation

$$[\Delta + n^2(\mathbf{r})k^2] \psi(\mathbf{r}) = 0, \qquad (3.1)$$

where $\Delta \equiv \partial_x^2 + \partial_y^2$, position vector $\mathbf{r} = (x, y) = (r \cos\theta, r \sin\theta)$, and the refractive index $n(\mathbf{r})$ is given by

$$n(\mathbf{r}) = \begin{cases} n_0 \left|\frac{d\zeta}{d\eta}\right|^{-1}, & \mathbf{r} \in \Omega_{in} \\ 1, & \mathbf{r} \in \Omega_{ex}. \end{cases} \qquad (3.2)$$

For asymptotically large $r$, the resonant mode function $\psi(\mathbf{r})$ must satisfy the outgoing-wave boundary condition,

$$\psi(\mathbf{r}) \sim h_k(\theta) \frac{e^{ikr}}{\sqrt{r}}, \qquad (3.3)$$

where $h_k(\theta)$ is the far-field angular distribution of the radiation emission. The outgoing-wave boundary condition leads to solutions exponentially decaying in time with discrete complex eigenvalues $k$ with $\text{Im}(k) < 0$. So their angular frequency $\omega$ becomes complex. The lifetime $\tau$ of these so-called 'resonant modes' or 'resonances' is given by the imaginary part of the angular frequency as $\tau = -1/[2\,\text{Im}(\omega)]$; the quality factor $Q$ of each resonance is defined by $Q = 2\pi\,\tau/T = -\text{Re}(\omega)/[2\,\text{Im}(\omega)]$ with the oscillation period of light wave $T = 2\pi/\text{Re}(\omega)$. The complex wave function $\psi$ equals $E_z$ and $H_z$ for TM and TE polarizations, respectively. The continuity relations at the cavity–air interface are given by

$$\psi_{in} = \psi_{ex}, \qquad \partial_\perp \psi_{in} = \partial_\perp \psi_{ex} \qquad \text{for TM modes,} \tag{3.4a}$$

$$\psi_{in} = \psi_{ex}, \qquad \frac{\partial_\perp \psi_{in}}{n_{in}^2} = \partial_\perp \psi_{ex} \qquad \text{for TE modes,} \tag{3.4b}$$

where $\psi_{in(ex)}$ and $\partial_\perp \psi_{in(ex)}$ are wave functions and their normal derivatives from the interior (exterior) of the cavity, respectively, and $n_{in}(= n_0|d\zeta/d\eta|^{-1})$ is the inhomogeneous refractive index from the interior region of the cavity evaluated at the boundary [32].

In BEM, the above scalar wave equation is replaced by corresponding BIEs by using Green's function and Green's second identity, and then the cavity boundary $\Gamma$ is discretized. In our case, the Green's function for the interior region in physical space is not known since the refractive index is spatially varying there, so we cannot obtain the BIE from the interior region of the cavity, while for the exterior region of the uniform refractive index the BIE of the conventional BEM can be used. Green's function in the exterior region is defined as the solution of

$$(\Delta + k^2)\,G^{ex}(\mathbf{r}, \mathbf{r}'; k) = \delta(\mathbf{r} - \mathbf{r}'), \tag{3.5}$$

where $\delta(\mathbf{r} - \mathbf{r}')$ is the Dirac $\delta$-function, and $\mathbf{r}$, $\mathbf{r}'$ are arbitrary points in the exterior region of the cavity. Green's function $G^{ex}(\mathbf{r}, \mathbf{r}'; k)$ for the exterior region is given by the zeroth order Hankel function of the first kind,

$$G^{ex}(\mathbf{r}, \mathbf{r}'; k) = -\frac{i}{4} H_0^{(1)}(k|\mathbf{r} - \mathbf{r}'|). \tag{3.6}$$

By multiplying Eq. (3.1) by Green's function, subtracting the resulting equation from the product of Eq. (3.5) and $\psi(\mathbf{r})$, and applying Green's second identity, we obtain the resulting boundary integral equation (BIE) as follows [17]

$$\psi(\mathbf{r}') = \oint_\Gamma \bigl(\psi(\mathbf{r})\,\partial_\perp G^{ex}(\mathbf{r}, \mathbf{r}'; k) - G^{ex}(\mathbf{r}, \mathbf{r}'; k)\,\partial_\perp \psi(\mathbf{r})\bigr) ds, \qquad \mathbf{r}' \in \Omega_{ex}, \tag{3.7a}$$

$$\frac{1}{2}\psi(\mathbf{r}') = PV \int_\Gamma \bigl(\psi(\mathbf{r})\,\partial_\perp G^{ex}(\mathbf{r}, \mathbf{r}'; k) - G^{ex}(\mathbf{r}, \mathbf{r}'; k)\,\partial_\perp \psi(\mathbf{r})\bigr) ds, \qquad \mathbf{r}' \in \Gamma, \tag{3.7b}$$

where $\partial_\perp$ is the normal derivative at point $\mathbf{r}$ on the boundary curve $\Gamma$ of region $\Omega_{ex}$ and $s \equiv s(\mathbf{r})$, which is the arc length along $\Gamma$ at $\mathbf{r}$ in the physical space; $PV$ means Cauchy principal value integration. In order to build the BIE for the interior region of cavity, we introduce the RV space which is obtained from the physical space by the inverse conformal mapping, $\eta = f^{-1}(\zeta)$, as shown in Figure 1 (c) and (d). In general, the inverse conformal mapping is not a one-to-one mapping and therefore the RV space corresponding to the OV space should be selected. In the RV space, functions, vectors, differential operators, and other relevant symbols will be expressed with tildes. Under the inverse conformal mapping, the scalar wave Eq. (3.1) transforms to

$$[\tilde{\Delta} + \tilde{n}^2(\tilde{\mathbf{r}})k^2]\,\tilde{\psi}(\tilde{\mathbf{r}}) = 0, \tag{3.8}$$

where $\tilde{\Delta} \equiv \tilde{\partial}_u^2 + \tilde{\partial}_v^2$, position vector $\tilde{\mathbf{r}} = (u, v)$, and refractive index $\tilde{n}(\tilde{\mathbf{r}})$ is given by

$$\tilde{n}(\tilde{\mathbf{r}}) = \begin{cases} n_0, & \tilde{\mathbf{r}} \in \tilde{\Omega}_{in} \\ \left|\dfrac{d\eta}{d\zeta}\right|^{-1}, & \tilde{\mathbf{r}} \in \tilde{\Omega}_{ex}. \end{cases} \tag{3.9}$$

We note that the cavity in the RV space has a constant refractive index profile as shown in Figure 1 (d). Thus, for the interior region of the cavity in the RV space, we can build the BIE,

$$\oint_{\tilde{\Gamma}} \Bigl(\bigl(2\,\tilde{\partial}_\perp G^{in}(\tilde{\mathbf{r}}, \tilde{\mathbf{r}}'; k) - \delta(\tilde{\mathbf{r}} - \tilde{\mathbf{r}}')\bigr)\tilde{\psi}(\tilde{\mathbf{r}}) - 2 G^{in}(\tilde{\mathbf{r}}, \tilde{\mathbf{r}}'; k)\,\tilde{\partial}_\perp \tilde{\psi}(\tilde{\mathbf{r}})\Bigr) d\tilde{s} = 0, \quad \tilde{\mathbf{r}}, \tilde{\mathbf{r}}' \in \tilde{\Gamma} \tag{3.10}$$

where $\tilde{\partial}_\perp \equiv \tilde{\mathbf{p}}(\tilde{\mathbf{r}}) \cdot \tilde{\nabla}|_{\tilde{\mathbf{r}}}$ and $\tilde{\mathbf{p}}(\tilde{\mathbf{r}})$ is the outward normal unit vector to the boundary curve $\tilde{\Gamma}$ of the region $\tilde{\Omega}_{in}$ at point $\tilde{\mathbf{r}}$; and $\tilde{s} \equiv \tilde{s}(\tilde{\mathbf{r}})$ is the arc length along $\tilde{\Gamma}$ at $\tilde{\mathbf{r}}$. The Green's function $G^{in}(\tilde{\mathbf{r}}, \tilde{\mathbf{r}}'; k)$ on the boundary of the cavity in the RV space is given by the zeroth order Hankel function of the first kind,

$$G^{in}(\tilde{\mathbf{r}}, \tilde{\mathbf{r}}'; k) = -\frac{i}{4} H_0^{(1)}(n_0 k |\tilde{\mathbf{r}} - \tilde{\mathbf{r}}'|), \tag{3.11}$$

and $\tilde{\partial}_\perp G^{in}(\tilde{\mathbf{r}}, \tilde{\mathbf{r}}'; k)$ is its normal derivative expressed as

$$\tilde{\partial}_\perp G^{in}(\tilde{\mathbf{r}}, \tilde{\mathbf{r}}'; k) = -\frac{i n_0 k}{4} \tilde{\mathbf{p}}(\tilde{\mathbf{r}}) \cdot \frac{\tilde{\mathbf{r}} - \tilde{\mathbf{r}}'}{|\tilde{\mathbf{r}} - \tilde{\mathbf{r}}'|} H_1^{(1)}(n_0 k |\tilde{\mathbf{r}} - \tilde{\mathbf{r}}'|), \tag{3.12}$$

where $H_1^{(1)}$ is the first order Hankel function of the first kind. To solve Eq. (3.10) numerically, the boundary must be discretized by dividing it into small boundary elements. The BIE (Eq. (3.10)) can be approximated as a discretized sum

$$\sum_m (\tilde{B}_{lm} \tilde{\partial}_\perp \tilde{\psi}_m + \tilde{C}_{lm} \tilde{\psi}_m) = 0, \tag{3.13}$$

where $\tilde{B}_{lm} \equiv \int_m \tilde{B}(\tilde{s}_l, \tilde{s}) d\tilde{s} = \int_m (-2 G^{in}(\tilde{s}, \tilde{s}_l; k)) d\tilde{s}$, $\tilde{C}_{lm} \equiv \int_m \tilde{C}(\tilde{s}_l, \tilde{s}) d\tilde{s} = \int_m (2\tilde{\partial}_\perp G^{in}(\tilde{s}, \tilde{s}_l; k) - \delta(\tilde{s} - \tilde{s}_l)) d\tilde{s}$, $\tilde{\partial}_\perp \tilde{\psi}_m \equiv \tilde{\partial}_\perp \tilde{\psi}(\tilde{s}_m)$, $\tilde{\psi}_m \equiv \tilde{\psi}(\tilde{s}_m)$, and $\int_m$ denotes integration over a boundary element with a midpoint $\tilde{s}_m$.

For the exterior region of the cavity, from Eq. (3.7b) we can write the following BIE,

$$\oint_\Gamma \left( (2\partial_\perp G^{ex}(\mathbf{r}, \mathbf{r}'; k) - \delta(\mathbf{r} - \mathbf{r}')) \psi(\mathbf{r}) - 2 G^{ex}(\mathbf{r}, \mathbf{r}'; k) \partial_\perp \psi(\mathbf{r}) \right) ds = 0, \quad \mathbf{r}, \mathbf{r}' \in \Gamma. \tag{3.14}$$

Equation (3.14) can be discretized as follows,

$$\sum_m (B_{lm} \partial_\perp \psi_m + C_{lm} \psi_m) = 0, \tag{3.15}$$

where $B_{lm} \equiv \int_m B(s_l, s) ds = \int_m (-2 G^{ex}(s, s_l; k)) ds$, $C_{lm} \equiv \int_m C(s_l, s) ds = \int_m (2\partial_\perp G^{ex}(s, s_l; k) - \delta(s - s_l)) ds$, $\partial_\perp \psi_m \equiv \partial_\perp \psi(s_m)$, and $\psi_m \equiv \psi(s_m)$. The size of the small boundary elements $ds$ in the physical space is a function of $d\tilde{s}$ in the RV space and this function can be derived from the conformal mapping $\zeta = f(\eta)$. For the case of TM resonant modes where both the wave function $\psi$ and its normal derivative $\partial_\perp \psi$ are continuous across the boundary, Eq. (3.13) and (3.15) can be written in a matrix form,

$$\begin{pmatrix} \tilde{B}_{lm} & \tilde{C}_{lm} \\ B_{lm} \left|\frac{d\zeta}{d\eta}\right|^{-1} & C_{lm} \end{pmatrix} \begin{pmatrix} \tilde{\partial}_\perp \tilde{\psi}_m \\ \tilde{\psi}_m \end{pmatrix} = 0. \tag{3.16}$$

In the above equation, we used the relation, $\psi_m = \tilde{\psi}_m$, which comes from the fact that electric field wave functions in the physical space and their counterpart functions in the RV space are the same as mentioned in section 2 and the normal derivative relation, $\partial_\perp \psi_m = \left|\frac{d\zeta}{d\eta}\right|^{-1} \tilde{\partial}_\perp \tilde{\psi}_m$, where $\left|\frac{d\zeta}{d\eta}\right|^{-1}$ is evaluated at the midpoint $\tilde{s}_m$; details of derivation of these relations can be found in Ref. [33]. Thus, for the case of TM resonant modes, the $2N \times 2N$ matrix $M(k)$ is defined by

$$M(k) = \begin{pmatrix} \tilde{B}_{lm} & \tilde{C}_{lm} \\ B_{lm} \left|\frac{d\zeta}{d\eta}\right|^{-1} & C_{lm} \end{pmatrix}, \tag{3.17}$$

where $N$ is the number of boundary elements. The resonant wave numbers $k_{res}$ are determined from the condition that the matrix $M(k)$ becomes singular, i.e.,

$$\det[M(k)] = 0. \tag{3.18}$$

For the case of TE resonant modes where the wave function $\tilde{\psi}$ and $\tilde{\partial}_\perp \tilde{\psi}/\tilde{n}^2$ are continuous across the boundary in the RV space, the above matrix $M(k)$ should be replaced with

$$M(k) = \begin{pmatrix} n_0^2 \tilde{B}_{lm} & \tilde{C}_{lm} \\ B_{lm} \left|\frac{d\eta}{d\zeta}\right|^{-1} & C_{lm} \end{pmatrix}. \tag{3.19}$$

Finally, the wave functions in the interior and exterior regions of cavities can be obtained from following BIEs (see Eq. (3.7a)) as follows

$$\tilde{\psi}_{in}(\tilde{\mathbf{r}}') = \oint_\Gamma \left( \tilde{\psi}(\tilde{s}) \, \tilde{\partial}_\perp G^{in}(\tilde{s}, \tilde{\mathbf{r}}'; k_{res}) - G^{in}(\tilde{s}, \tilde{\mathbf{r}}'; k_{res}) \, \tilde{\partial}_\perp \tilde{\psi}(\tilde{s}) \right) d\tilde{s}, \qquad \tilde{\mathbf{r}}' \in \tilde{\Omega}_{in} \tag{3.20a}$$

$$\psi_{ex}(\mathbf{r}') = \oint_\Gamma \left( \psi(s) \, \partial_\perp G^{ex}(s, \mathbf{r}'; k_{res}) - G^{ex}(s, \mathbf{r}'; k_{res}) \, \partial_\perp \psi(s) \right) ds, \qquad \mathbf{r}' \in \Omega_{ex}. \tag{3.20b}$$

The wave function $\psi_{in}(\mathbf{r}')$ in the interior region of the cavity in the physical space can be obtained from $\tilde{\psi}_{in}(\tilde{\mathbf{r}}')$ using the relation $\psi_{in}(\mathbf{r}') = \tilde{\psi}_{in}(\tilde{\mathbf{r}}')$, where $\mathbf{r}'$ is related to $\tilde{\mathbf{r}}'$ by the conformal mapping.

## 4. Numerical calculation in limaçon-shaped transformation cavities

The conformal transformation, which maps the unit circle to the limaçon [14, 34], is given by

$$\zeta = \beta(\eta + \epsilon\eta^2), \tag{4.1}$$

where $\eta$ and $\zeta$ are complex variables that denote positions in the two complex planes, $\epsilon$ is a deformation parameter, and $\beta$ is a positive scaling factor. The refractive index $n(x, y)$ in the physical space derived from Eq. (3.2) is given by

$$n(x, y) = \begin{cases} \frac{n_0}{\beta|\sqrt{1 + 4\epsilon\zeta/\beta}|}, & (x, y) \in \Omega_{in} \\ 1, & (x, y) \in \Omega_{ex} \end{cases} \tag{4.2}$$

where $n_0$ is the constant refractive index of a circular cavity in the OV space. The refractive index $\tilde{n}(u, v)$ derived from Eq. (3.9) in the RV space is given by

$$\tilde{n}(u, v) = \begin{cases} n_0, & (u, v) \in \tilde{\Omega}_{in} \\ \beta|1 + 2\epsilon\eta|, & (u, v) \in \tilde{\Omega}_{ex}. \end{cases} \tag{4.3}$$

Figure 1 (c) and (d) show a limaçon-shaped TC and an inversely transformed circular cavity obtained from Eq. (4.1) with $\epsilon = 0.24$, $n_0 = 2.0$, and $\beta = 1.0$ in the physical space and the RV space, respectively. We note that the refractive index $\tilde{n}(u, v)$ given by the inverse conformal transformation of Eq. (4.1) is a double-valued function. The proper branch of refractive index $\tilde{n}(u, v)$ must be chosen so that the circular cavity in the RV space is same as the circular cavity in the OV space.

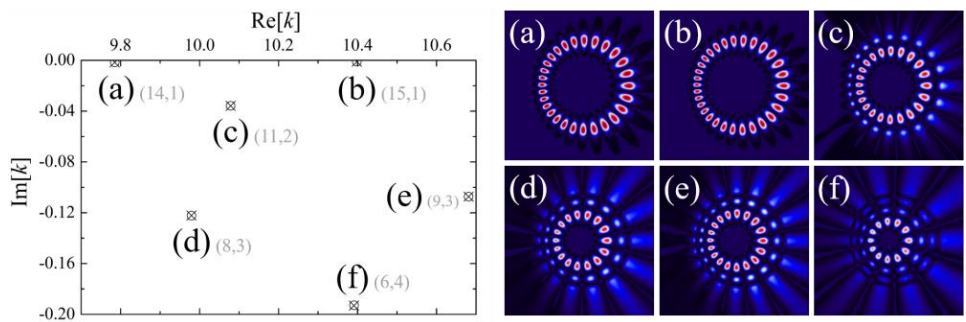

**Figure 2.** Resonance positions in complex $k$-space and near-field intensity patterns for TM modes found by BEM in the limaçon-shaped TC with $\epsilon = 0.15$, $n_0 = 1.8$, and $\beta = 0.769$. Crosses and open circles mark the positions of a pair of nearly degenerate resonant modes labeled by same $(m, l)$ for even- and odd-parity modes with respect to $x$-axis on the left. We depicted only the even- parity mode patterns on the right.

Using the BEM formulated in the previous section, we obtained TM resonances when $\epsilon = 0.15$, $n_0 = 1.8$, and $\beta = 0.769$. The positive scaling factor $\beta$ is introduced to obtain anisotropic WGMs supported by TIR which is called conformal Whispering Gallery Mode (cWGM) [14]. The condition for TIR in generic TCs is given by $|d\zeta/d\eta|^{-1} \geq 1$; in this case, the condition becomes $\beta \leq \beta_{max} = 1/(1 + 2\epsilon)$ and we use $\beta = \beta_{max}$ which is the minimal condition for TIR. The resonant wave numbers $k_{res}$ can be obtained from $\det[M(k)] = 0$ through the root-finding algorithm such as the Newton-Raphson method. All of resonances in the range of $9.7 < \text{Re}[k] < 10.7$ and $0 < \text{Im}[k] < -0.2$ are marked in the complex $k$-space as shown in Figure 2. Since each mode has correspondence with the resonances in a uniform index circular cavity in the OV space which are labeled by the azimuthal mode number $m$ and the radial mode number $l$, we simply denote the modes by $(m, l)$.

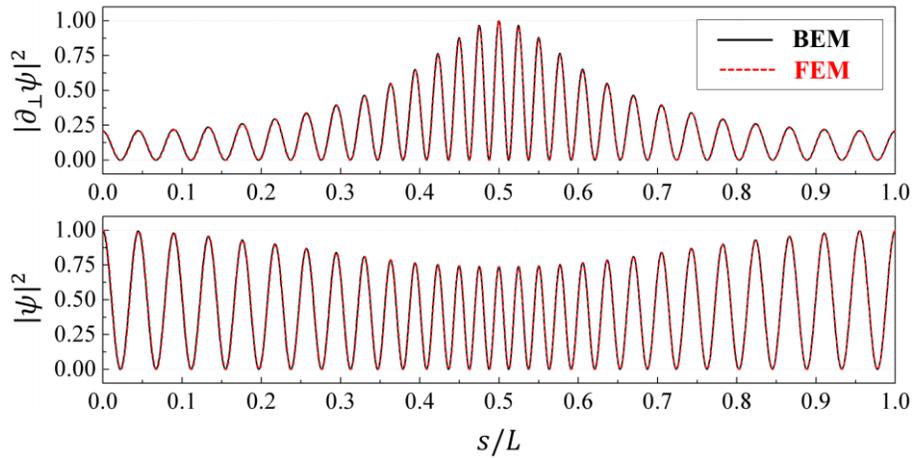

**Figure 3.** Comparison of the normalized $|\partial_\perp \psi|^2$ and $|\psi|^2$ along the normalized arc length $s/L$ for the $(14, 1)$ TM cWGM in a limaçon-shaped TC with $\epsilon = 0.15$, $n_0 = 1.8$, and $\beta = 0.769$ obtained by BEM and FEM, respectively.

In order to confirm the validity of our BEM, we compared the results of BEM with that of the corresponding simulation performed by COMSOL Multiphysics v.5.3, a commercial FEM-based electromagnetic solver. In Figure 3, we depict the normalized values of $|\partial_\perp \psi|^2$ and $|\psi|^2$ along the arc length $s$ normalized with total arc length $L$ for a resonant mode, (14,1) with the complex wave number $k_{res} = 9.785 - i\,0.00158$ and one can see that the two results match almost exactly. The (14,1) cWGM has a high-$Q$ factor ($Q \approx 3097$) because of the evanescent leakage by TIR.

We plotted the near-field wave patterns and the far-field intensity distributions of the cWGM obtained by the two methods in Figure 4. The far-field distribution exhibits bidirectionality due to the tunneling emission at the rightmost position of the limaçon shaped TC where the refractive index is lowest. In Figure 4(a), the top, bottom-left, and bottom-right patterns are the intensity, the real part, and the imaginary part of the complex wave functions obtained from BEM, respectively. Figure 4(b) are the results for the same mode obtained from FEM. As shown in Figure 4(c), the far-field intensity distribution obtained with BEM also agree well with the result obtained with FEM. Incidentally, the BEM based on the Green's function can obtain the wave function value at any spatial point using only $\partial_\perp \psi$ and $\psi$ at the cavity boundary and it is one of the notable advantages of this method.

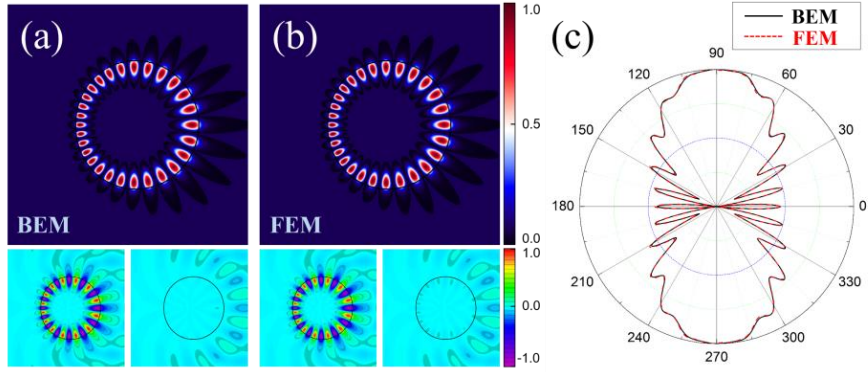

**Figure 4.** Comparisons between (a), (b) near-field, and (c) far-field patterns obtained by BEM and FEM for even-parity $(14,1)$ TM cWGM in a limaçon-shaped TC with $\epsilon = 0.15$, $n_0 = 1.8$, and $\beta = 0.769$. Top, bottom-left, and bottom-right images in (a) and (b) are $|\psi|^2$, $\text{Re}(\psi)$, and $\text{Im}(\psi)$, respectively. In (c), the solid black and the dashed red lines are the distributions obtained by BEM and FEM, respectively.

In our case, all resonances except those with $m = 0$ are exist in nearly degenerate pairs due to the breaking of rotational symmetry. The mode pairs are formed with even- and odd-parity with respect to the mirror symmetry axis ($x$-axis) and are very close to each other in the complex $k$-space as shown in Figure 2. In Figure 5, we depicted the mode intensity patterns and the corresponding far-field intensity distributions of a nearly degenerate $(14,1)$ cWGM pair. The even-parity (Figure 5 (a)) and the odd-parity mode (Figure 5 (b)) with respect to $x$-axis have $k_{res} = 9.785240667 - i\, 0.0015797513$ and $k_{res} = 9.785240670 - i\, 0.0015797508$, respectively. The far-field intensity distributions of the cWGM pair show out of phase interference patterns but all exhibit bidirectional emission feature as shown in Figure 5 (c).

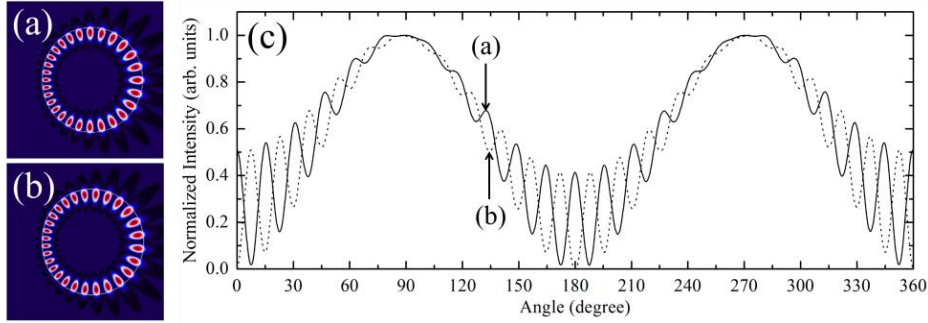

**Figure 5.** The nearly degenerate pair of $(14,1)$ TM cWGMs obtained by BEM in the limaçon-shaped TC with $\epsilon = 0.15$, $n_0 = 1.8$, and $\beta = 0.769$. (a) and (b) are wave intensity patterns of the pair and (c) are their far-field intensity distributions.

Using the BEM, we also obtained another resonant mode in a limaçon-shaped TC with $\epsilon = 0.24$, $n_0 = 2.0$, and $\beta = 1.0$. The complex wave number of the mode is $k_{res} = 11.913 - i\, 0.107$ and the intensity pattern of the resonant mode is shown in Figure 6 (a). This resonant mode has low-$Q$ factor ($Q \approx 56$) and the corresponding far-field intensity distribution is shown in Figure 6 (c). Contrary to the bidirectional far-field distribution of the above $(14,1)$ cWGM, this low-Q mode has a unidirectional far-field intensity distribution. From these results, one can note that the $Q$-factors and emission directionalities of the resonant modes in TCs depend on the scaling factor $\beta$ as well as the deformation parameter $\epsilon$ [35]. We also obtained the resonant mode under the same parameters by FEM and the resultant wave number of the mode is $k_{res} = 11.913 - i\, 0.108$, which agrees well with the BEM result. Also, the intensity pattern of the corresponding mode and the far-field intensity distribution obtained by FEM are shown in Figs. 6 (b) and (c), respectively, which nearly match the results of BEM as well.

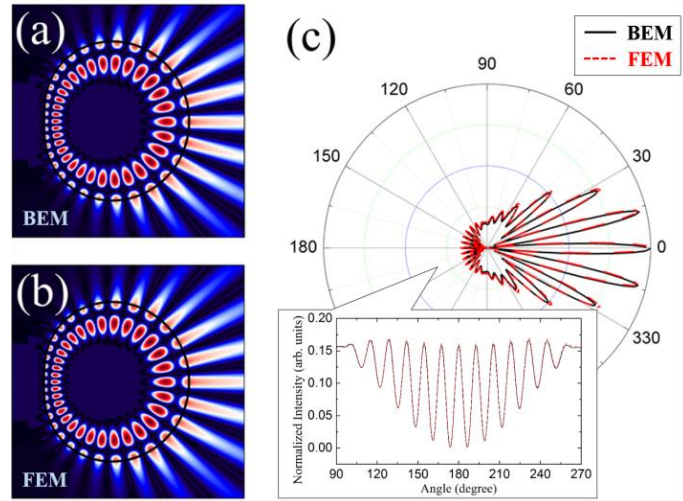

**Figure 6.** The even-parity $(16, 2)$ TM resonant mode in the limaçon-shaped TC with $\epsilon = 0.24$, $n_0 = 2.0$ and $\beta = 1.0$. (a) and (b) are wave intensity patterns obtained by BEM and FEM, respectively. (c) is far-field intensity distribution which are obtained by BEM (solid black lines) and FEM (dashed red lines). Inset of (c) is a zoom-in plot for the range of 90~270 degree.

## Summary

The cWGMs in TCs can simultaneously possess useful properties, such as an ultrahigh-$Q$ factor and directional light emission, which are seemingly incompatible. In order to obtain the resonant mode functions in TCs designed by conformal TO, we have developed a pure BEM based on the fact that the Green's functions of interior and exterior regions are known in the RV and the physical spaces, respectively, which are connected by a conformal mapping. For the verification of our BEM, we have calculated resonant modes in limaçon-shaped TCs and compared those with the corresponding results from FEM. Complex wave numbers, mode patterns, and far-field intensity distributions of the resonant modes obtained by the BEM almost match those obtained by FEM.

Like the conventional BEM for homogeneous dielectric cavities, the newly proposed BEM has advantages in computing resonances in the TCs, e.g. a relatively simple formalism and efficiency, especially in finding highly-excited states. It can also be used for a measure of reliability for FEM calculations in non-piecewise-constant media. We expect that our method will be extended to the calculation of resonant modes in TCs with more complex geometries.

### Acknowledgements

This research at KNU (J.C., S.-Y.L., S.-J.P., S.R., M.C. & I.K.) was supported by the National Research Foundation of Korea(NRF) grant funded by the Korean government (MSIT) (No. 2017R1A2B4012045 and No. 2017R1A4A1015565). J.-W.R. was supported by the Institute for Basic Science in Korea (IBS-R024-D1).

### Author Contributions

S.-Y.L. conceived the original idea. J.-W.R., S.-Y.L., M.C. and I.K. formulated the method. J.-W.R., J.C., Y.K. and S.-J.P. performed numerical calculation. J.-W.R., J.C., S.R., M.C. and I.K. wrote the manuscript. All authors analyzed the data, discussed the results and provided feedback for the manuscript.

### Additional Information

**Competing Interests:** The authors declare no competing interests.